\documentclass[11pt,twoside]{article}

\usepackage{epsfig,wrapfig}

\advance \textwidth by -5cm
\advance \textheight by -6cm
\def\e{\begin{equation}}
\def\f{\end{equation}}
\def\_#1{{\bf #1}}

\def\.{\cdot}

\begin{document}

\title{\bf{Balanced right/left-handed mixtures
of quasi-planar chiral inclusions}}




\def\affil#1#2{\begin{itemize} \item[$^1$] #1 \item[$^2$] #2 \end{itemize}}

\author{\underline{R. Marqu\'es}$^{1a}$, F. Mesa$^2$, L. Jelinek$^{1b}$
and J.D. Baena$^{1c}$}

\maketitle

\affil{Dept. of Electronics and Electromagnetism, University of
Seville (Spain), \\ e-mail: (a) marques@us.es, (b) bish@atlas.cz,
(c) juan$_-$dbd@us.es}{Dept. of Applied Physics 1, University of
Seville (Spain), e-mail: mesa@us.es}

\thispagestyle{empty}

\begin{abstract}
\noindent Some novel quasi-planar chiral inclusions, feasible from
standard photo-etching techniques, are proposed. It is shown that
such inclusions can be designed in order to present balanced
electric, magnetic and magneto-electric polarizabilities. Using
these inclusions, random and periodic bi-isotropic artificial
metamaterials exhibiting a balanced positive/negative refractive
index can be build up. These metamaterials would exhibit
reasonable bandwidths and excellent matching to free space
\end{abstract}

\subsection*{1.~Introduction}

In spite of some proposal in such direction \cite{Martin} --
\cite{Alitalo-2}, the development of bulk isotropic metamaterials
exhibiting negative refractive index (NRI) and good matching to
free space is still a challenging issue. A promising approach to
this problem could take advantage of the simultaneous electric and
magnetic polarizability of chiral inclusions in order to obtain a
mixture with simultaneously negative $\varepsilon$ and $\mu$. As
far as we know, the first proposal in such direction was made in
\cite{Tretyakov}, and further developed in
\cite{Tretyakov-Shivola}. In these works racemic and chiral
mixtures of chiral metallic inclusions were proposed as a
practical way of designing left-handed metamaterials. Other
proposals taking advantage of chirality for negative refractive
artificial media design have also been made (see \cite{Monzon} and
\cite{Pendry-chiral}, for instance). In this contribution we will
further develop this approach presenting some novel NRI
bi-isotropic metamaterial designs that exhibit reasonable
bandwidths and excellent matching to free space.

\subsection*{2.~Balanced positive/negative refractive index (BPNRI)
metamaterials}

BPNRI metamaterials are defined as bi-isotropic media exhibiting
balanced electric, $\chi_e$, and magnetic $\chi_m$
susceptibilities, that is
\begin{equation} \chi_e(\omega) = \chi_m(\omega)\,,
\label{balanced-media}\end{equation} over a wide bandwidth which
includes both positive and negative values of
$\varepsilon_r=1+\chi_e$ and $\mu_r=1+\chi_m$. Such media are
characterized by the following properties:
 \begin{itemize}
 \item Wide NRI pass-band for the positive and/or the negative
 circularly polarized TEM eigenwaves, defined by the condition
 $\sqrt{\varepsilon_r\mu_r}\pm\kappa <0$ \cite{Mackay} where the negative
 sign for the square root must be chosen if $\varepsilon$ and $\mu$ are both
 negative.
 \item No forbidden bands, as it is deduced from the dispersion equation $k^{\pm}
 = k_0\left(\sqrt{\mu_r\varepsilon_r} \pm\kappa\right)$ with $k_0=
 \omega\sqrt{\epsilon_0\mu_0}$.
  \item Transition between the NRI and PRI pass-bands through a
 zero phase velocity point (a behavior similar to that previously
 reported for some transmission line metamaterials \cite{Caloz}).
 \item Good matching (perfect matching for paraxial rays) to free
 space, due to the matching of TEM impedances
 $\eta=\sqrt{\mu/\varepsilon}=\eta_0$.
 \end{itemize}

In the following we will develop a design with $|\kappa| \sim
\chi_e, \chi_m $. In such case only one of the TEM circularly
polarized eigenwaves will exhibit negative refraction and the NRI
pass-band is defined by the condition $\chi_e=\chi_m < -0.5$
\cite{Marques-MOTL}

\subsection*{3. Balanced quasi-planar chiral inclusions}

Figure 1 shows two quasi-planar chiral inclusions suitable for
balanced metamaterial design: the chiral split ring resonator
(Ch-SRR) and the chiral spiral resonator (Ch-SR). They are the
broadside-coupled versions of the spiral resonator
\cite{Baena-spiral} and the NB-SRR \cite{Baena-circuits}
previously proposed by some of the authors. In both cases the
frequency of resonance con be obtained from an $LC$ circuit model:
$\omega_0=1/\sqrt{LC}$ where $L$ and $C$ are the effective
inductance and capacitance of the inclusion (analytical
expressions can be found in \cite{Marques-AP}). The
polarizabilities can be obtained following the standard technique
developed in \cite{Marques-AP} and in \cite{Baena-spiral} --
\cite{Baena-circuits}. This calculation gives: \begin{eqnarray}
     \label{alphamm}
     \alpha_{zz}^{mm} &=& \alpha_0^{mm}
    \frac{\omega^2}{\omega_0^2-\omega^2+j \omega R/L}
    \;;\;\;\;\;\;\;\;\;\;\;\;\;\;\;\;\;\;\;\;\;\;\;
    \alpha_0^{mm} = \frac{\pi^2r^4}{L}
      \\
      \label{alphaem}
     \alpha_{zz}^{em} &=& \pm \,
     \alpha_0^{em}\,
     \left(\frac{\omega_0}{\omega}\right)\,
     \frac{\omega^2}{\omega_0^2-\omega^2+j \omega R/L}
     \;;\;\;\;\;\;\;\;\,
    \alpha_0^{em} = j\frac{n\pi r^2
    t}{\omega_0L}\left(\frac{C_0}{C}\right)
      \\
     \label{alphaee}
     \alpha_{zz}^{ee} &=& \alpha_0^{ee}
     \left(\frac{\omega_0}{\omega}\right)^2
      \frac{\omega^2}{\omega_0^2-\omega^2+j \omega R/L}
      \;;\;\;\;\;\;\;\;\;\;\;\;\;\;
    \alpha_0^{ee} = \frac{(nt)^2}{\omega_0^2L}\left(\frac{C_0}{C}\right)^2 \\
    \alpha_{xx}^{ee} &=& \alpha_{yy}^{ee} = \alpha_0
    \;;\;\;\;\;\;\;\;\;\;\;\;\;\; \;\;\;\;\;\;\;\;\;\;\;\;\;
    \;\;\;\;\;\;\;\;\;\;\;\;\;\;\;\;\;\;\;\;
    \alpha_0 = \varepsilon_0\frac{16}{3}r_{ext} \label{alpha0}
    \,,
\end{eqnarray} where $n=1$ for the Ch-SR and $n=2$ for the Ch-SRR
(see \cite{Marques-AP} for more details on notation). From
(\ref{alphamm}) -- (\ref{alphaee}) it follows that
\begin{equation}\label{alpha-prop}
\alpha_{zz}^{mm}\alpha_{zz}^{ee} + (\alpha_{zz}^{em})^2=0 \;,
\end{equation} which is a general property arising from the $LC$
nature of the model \cite{Tretyakov-Shivola}. In order to obtain a
balanced design we will further impose \begin{equation}
\alpha_0^{ee} = \mu_0\varepsilon_0\alpha_0^{mm} \,,
\label{alpha-condition}\end{equation} which is satisfied provided
that
\begin{equation} t\lambda_0 = \frac{2}{n}\frac{C}{C_0}\,(\pi r)^2
\label{particle-condition}\end{equation} where $\lambda_0$ is the
wavelength at resonance. Incidentally, since the frequency of
resonance of the Ch-SRR is twice that of the Ch-SR,
(\ref{particle-condition}) provides exactly the same geometrical
parameters for both inclusions. The accuracy of this expression
for giving a balanced design has been shown in \cite{Marques-MOTL}
by electromagnetic simulation.
\begin{figure}
\centering \epsfig{file=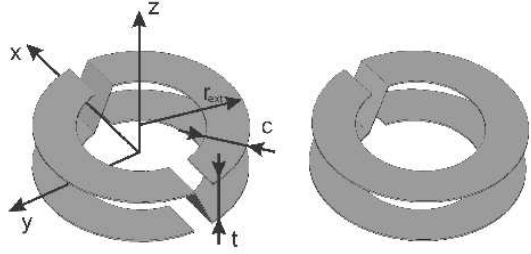, width=0.5\textwidth} \caption{
Two quasi-planar chiral inclusions, the chiral SRR (Ch-SRR, left)
and the chiral spiral resonator (Ch-SR, right). For practical
designs metallic rings can be photo-etched on both sides of a
dielectric board and connected through a via-hole.}
\end{figure}

\subsection*{4. Isotropic arrangement}

In order to obtain a BPNRI metamaterial the inclusions shown in
Fig.1 must be properly arranged. The main advantage of the Ch-SR
is its small electrical size (balanced designs with sizes of about
$\lambda_0/13$ can be obtained \cite{Marques-MOTL}). However, the
Ch-SR has not enough symmetry to allow for the design of an
isotropic cubic resonator, suitable for a periodic isotropic
design \cite{Baena}. This condition is fulfilled by the Ch-SRR
which allows for the design of cubic arrangements satisfying the
cubic T group of symmetry (in Schoenflies notation), which is
enough to guarantee an isotropic metamaterial design \cite{Baena}.
Therefore, Ch-SRs are suitable for the design of random BPNRI
media and the Ch-SRR is appropriate for the design of periodic
BPNRI media. In both cases a Lorentz homogenization procedure
provides the following relation between the macoscopic fields and
volume polarizations
\begin{equation} M = N
\left\{\mu_0\hat{\alpha}^{mm}\left(H+\frac{M}{3}\right) -
\hat{\alpha}^{em}\left(E+\frac{P}{3\varepsilon_0}\right)\right\}
\, \label{M1}\end{equation} \begin{equation} P = N
\left\{\hat{\alpha}^{ee}\left(E+\frac{P}{3\varepsilon_0}\right) +
\mu_0\hat{\alpha}^{em}\left(H+\frac{M}{3}\right)\right\} \,,
\label{P1}\end{equation} where $N$ is the number of particles per
unit volume, and $\hat{\alpha}^{mm}$, $\hat{\alpha}^{ee}$,
$\hat{\alpha}^{em}$ are some average polarizabilities given by
$\hat{\alpha}^{mm}=\alpha_{zz}^{mm}/3$,
$\hat{\alpha}^{em}=\alpha_{zz}^{em}/3$ and
$\hat{\alpha}^{ee}=(\alpha_{xx}^{ee}+\alpha_{yy}^{ee}+\alpha_{zz}^{ee})/3$

For a balanced design of the inclusions, taking
(\ref{particle-condition}) into account, we finally find
\begin{eqnarray} \chi_e &=& \frac{N}{3\Lambda}\left\{
\frac{2\alpha_0}{\varepsilon_0} \left(\frac{\omega_0^2}{\omega^2}
- 1\right) + \mu_0\alpha_0^{mm} \left(\frac{\omega_0^2}{\omega^2}
-\frac{2N\alpha_0}{9\varepsilon_0}\right) \right\} \label{chie3}
\\ \chi_m &=& \frac{N\mu_0\alpha_0^{mm}}{3\Lambda}\,\left(1 -
\frac{2N\alpha_0}{9\varepsilon_0}\right) \label{chim3} \\ \kappa
&=& \pm
\frac{\omega_0}{\omega}\,\frac{N\mu_0\alpha_0^{mm}}{3\Lambda}
\label{kappa3} \end{eqnarray} where \begin{equation} \Lambda = K
\left\{\frac{\omega_0^2}{\omega^2} -1
+\frac{N\mu_0\alpha_0^{mm}}{9}\left(1+\frac{\omega_0^2}{K\omega^2}\right)
+ j\frac{R}{\omega L} \right\} \;\;;\;\;\;\;\; K =
\left(1-\frac{2N\alpha_0}{9\varepsilon_0}\right) \label{Lambda}
\end{equation} These expresions satisfy (\ref{balanced-media}) in the
limit $\omega_0/\omega \rightarrow 1$. Since most resonant
metamaterials have a moderate bandwidth ($\sim 10\%$ or less),
this condition is approximately fulfilled inside the left-handed
pass-band.

Figure 2 shows the propagation constants and impedances, $k^{\pm}
= k_0\left(\sqrt{\mu_r\varepsilon_r} \pm\kappa\right)$ and
$\eta=\sqrt{\varepsilon/\mu}$, computed from (\ref{chie3}) --
(\ref{kappa3}) for the TEM eigenwaves in a compact arrangement of
balanced Ch-SRs randomly oriented. $N$ corresponds to a cubic
\emph{fcc} lattice of balls with radius $a=1.1 r_{ext}$ which
contains the Ch-SRs, i.e. $N=0.74\left(\frac{4}{3}\pi (1.1
r_{ext})^3\right)^{-1}$. The remaining parameters are given in the
caption. As it can be seen, a BPNRI behavior is quite
approximately obtained.
\begin{figure}
\centering \epsfig{file=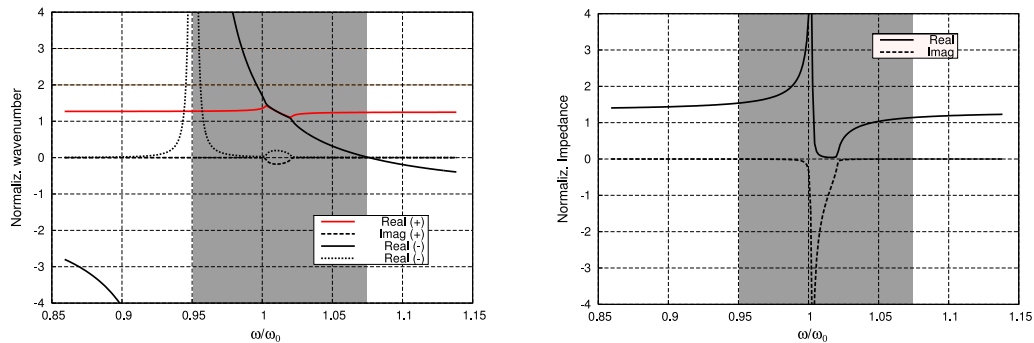, width=0.85\textwidth} \caption{
Plots of $k^\pm/k_0$ ($k_0=\omega\sqrt{\varepsilon_0\mu_0}$)
(left) and $\eta/\eta_0$ against $\omega/\omega_0$ (right) for a
cubic \emph{fcc} lattice of balanced Ch-SRs with $c/r_{ext}=0.2$,
and $t/r_{ext}=0.306$. Substrate is foam with
$\epsilon=\epsilon_0$ and $N$ is calculated in the text. The NRI
pass-band is marked in the figures.}
\end{figure}

\subsection*{5.~Conclusion}

Along this paper a fully analytical method for designing
bi-isotropic balanced positive/negative refractive index
metamaterials has been presented. Some new quasi-planar chiral
inclusions have been proposed for this purpose. These inclusions
can be easily manufactured by standard photo-etching techniques
and the condition for balanced design can be expressed in a quite
simple way (\ref{particle-condition}). Hopefully, the interesting
properties of bi-isotropic BPNRI metamaterials, such as the smooth
transition between the NRI and the PRI pass-bands and the
excellent matching to free space, could find application in
focusing devices and other applications.


{\small

\begin{thebibliography}{10}
\setlength{\itemsep}{-1ex}

\bibitem{Martin} P.Gay-Balmaz, O.J.F. Martin
\emph{J. App. Phys.} {\bf 92}, 2929 (2002).

\bibitem{Simowski} C. R. Simowski and S. He
\emph{Phys. Lett. A}, {\bf 311}, 254 (2003).

\bibitem{Koschny} Th. Koschny, L. Zhang, and C. M. Soukoulis,
Phys. Rev. B, {\bf 71}, 121103 (2005).

\bibitem{Holloway} C.L.Holloway, E.F.Kuester, J.Baker-Jarvis,
and P.A.Kabos, \emph{IEEE Trans. on Antennas and Propagation} {\bf
51}, 2596 (2003).

\bibitem{Vendik} I.Vendik, O.Vendik, I.Kolmakov, and M.Odit,
\emph{Opto-Electronics Review} {\bf 14}, 179 (2006).

\bibitem{Baena} J.D.Baena, L.Jelinek, R.Marqu\'es,
J.Zehentner, \emph{App. Phys. Let.}, {\bf 88}, 134108 (2006)

\bibitem{Hoefer} W.J.R.Hoefer, P.P.M. So, D.Thompson, and M.Tentzeris,
\emph{IEEE Int. Microwave Symposium Digest}, pp.313-316, Long
Beach (CA), USA (2005).

\bibitem{Grbic} A.Grbic and G.V.Eleftheriades, \emph{J. Appl. Phys.}
{\bf 98}, 043106 (2005).

\bibitem{Alitalo-2} P.Alitalo, S.Maslovski, and S.Tretyakov, \emph{J.
Appl. Phys.} {\bf 99}, 124910 (2006).

\bibitem{Tretyakov} S.A.Tretyakov \emph{Analytical modelling in
applied electromagnetics}, Artech House, Norwood MA, 2003.

\bibitem{Tretyakov-Shivola} S.A.Tretyakov, A.Sihvola, and L.Jylh
\emph {Photonics and Nanostruct. Fund. and Appl.} {\bf 3}, 107
(2005)

\bibitem{Monzon} C.Monzon, D.W.Forester \emph{Phys. Rev. Lett.},
{\bf 95}, 123904 (2005).

\bibitem{Pendry-chiral} J.B.Pendry \emph{Science}, {\bf 306},
1353 (2004).

\bibitem{Mackay} T.G.Mackay \emph{Microwave and Opt.
Tech. Lett.}, {\bf 45}, 120 (2005).

\bibitem{Caloz} C. Caloz and T. Itoh, \emph {Proc. of the IEEE-MTT
Int�l Symp}, vol.1 Philadelphia, PA, pp.195 (2003).

\bibitem{Marques-MOTL} R.Marqu\'es, L.Jelinek, and F.Mesa
\emph{arxiv:physics/0610071v1} (2006).

\bibitem{Baena-spiral} J.D.Baena, R.Marqu\'es, F.Medina, J.Martel,
\emph{Phys. Rev. B}, vol. 69, paper 014402, 2004.

\bibitem{Baena-circuits} J.D. Baena, J.Bonache, F.Mart\'in, R.Marqu\'es,
F.Falcone, T.Lopetegi, M.A.G.Laso, J.Garc�a, I.Gil and M.Sorolla,
\emph{IEEE Trans. on Microwave Theory and Tech.} {\bf 53}, 1451
(2005).

\bibitem{Marques-AP} R.Marqu\'es, F.Mesa, J.Martel and F.Medina
\emph{IEEE Trans. Ant. Propag.} {\bf 51}, 2572 (2003).

\end{thebibliography}
}
\end{document}